\documentclass[10pt]{iopart}
\usepackage{graphicx}
\usepackage{dcolumn}
\begin{document}
\title[]
%[Equation of state for liquids and glasses]
{
Equilibrium and out of equilibrium thermodynamics
in supercooled liquids and glasses
}
\author{S. Mossa\dag\ddag, E. La Nave\ddag, P. Tartaglia\ddag and F. Sciortino\ddag}
\address{\dag Laboratoire de Physique Th\'eorique des Liquides,
Universit\'e Pierre et Marie Curie , 4 Place Jussieu,
75252 Paris C\'edex 05, France}
\address{\ddag Dipartimento di Fisica, INFM Udr and
Center for Statistical Mechanics and Complexity,
Universit\`a di Roma "La Sapienza",
Piazzale Aldo Moro 2, I-00185, Roma, Italy}
\begin{abstract}
We review the inherent structure thermodynamical formalism and the
formulation of an equation of state for liquids in equilibrium
based on the (volume) derivatives of the statistical properties of
the potential energy surface. We also show that, under the 
hypothesis that during aging the system explores states associated
to equilibrium configurations, it is possible to generalize the
proposed equation of state to out-of-equilibrium conditions. The
proposed formulation is based on the introduction of one additional 
parameter which, in the chosen thermodynamic formalism, can be chosen 
as the local minima where the slowly relaxing out-of-equilibrium 
liquid is trapped.
\end{abstract}
\section{Introduction}
The possibility of a consistent description of the thermodynamics
of equilibrium and out-of-equilibrium (glass) supercooled liquids
has been and it is an important research
line~\cite{davies,kurchan,teoneowanoezer,fdprl,kob2000,
scala,kobbarrat, dileonardo2000,lanave02,aging02}. 
In recent years, the inherent structure (IS) formalism by Stillinger and
Weber~\cite{sw} 
%, boosted by the increase of computational capabilities, 
has significantly contributed to the understanding
of the physics of supercooled liquids and appears to offer a
powerful and simple approach for developing a thermodynamics of
out-of-equilibrium (OOE) states. Indeed, on one side, the IS formalism
provides a transparent way for writing the partition function in
terms of the ``base'' of the local minima (inherent structures) of
the underlying potential energy landscape (PEL). On the other
side, state of the art computer simulations 
%---mainly atomistic molecular  dynamics (MD) and Monte Carlo---
provide the possibility of a statistically complete sampling of
the PEL explored in equilibrium conditions in a wide temperature
range.  The numerical analysis of configurations extracted from
the canonical ensemble allows us to calculate the energy depth of
the ISs explored during the dynamical evolution, to characterize
the volume of their basins of attraction, and to give
estimates of their degeneracy (configurational entropy).
Eventually it allows us to directly estimate the free energy of
the system in terms of landscape properties.

Recently~\cite{lanave02} it has been shown that
%, building on the IS formalism, 
it is possible to write down a model equation of
state (EOS)~\cite{deben} expressed only in terms of quantities
describing the statistical properties of the PEL.  The crucial
step in this process is the evaluation/modellization of the volume
dependence of the number of basins, of their energy distribution
and volume. The landscape based PEL-EOS is able to predict the
thermodynamics of the system in equilibrium~\cite{lanave02} and,
even more interestingly, in out-of-equilibrium~\cite{aging02}---
when the aging system explores states related to equilibrium
configurations.

Although the calculations we have explicitely performed deal with
the simulation of the Lewis and Wanstr\"{o}m (LW) model for the
fragile molecular liquid orthoterphenyl
(OTP)~\cite{mossa02,lewis}, our results are general. Here we
review and discuss 
some general implications
of our results on the understanding of the thermodynamics of
supercooled liquids and glasses.
\section{The constant volume free energy}
In pioneering papers~\cite{sw}, Stillinger and Weber have shown
that the partition function ${\cal Z}(T)$ at constant volume $V$
 can be written as
\begin{equation}
{\cal Z}(T)=\int d e_{IS} \;\Omega(e_{IS}) \;e^{-\beta e_{IS}}e^{-\beta F_{vib}(e_{IS},T)},
\label{eq:partition}
\end{equation}
where  $\beta=1/k_BT$, $e_{IS}$ is the depth 
of the local potential energy minima
(IS) of the PEL, $\Omega(e_{IS}) de_{IS}$ is the number of
potential energy minima with energy between $e_{IS}$ and
$e_{IS}+de_{IS}$, and $F_{vib}(e_{IS},T)$ describes the 
free energy of the system constrained in one of the basin of depth $e_{IS}$,
averaged over all basins of depth $e_{IS}$.
Following Stillinger and Weber~\cite{sw}, a basin is defined 
as the set of points in configuration space which lead to the same
local minima under a steepest descent path. The  power of this
formulation relies in the fact that the procedures used to
associate to each system configuration the corresponding IS are
operationally well-defined through constant volume minimization
techniques. Numerical evaluation of the density of states in
the local minima allows us to calculate the harmonic contribution to
the basin free energy. Starting from Eq.~(\ref{eq:partition}), the
free energy of the system can be written as
\begin{equation}
F(T) = \langle e_{IS}(T) \rangle - T S_{conf}(\langle e_{IS}(T) \rangle) +
F_{vib} (T,\langle e_{IS}(T) \rangle);
\label{eq:free_energy1}
\end{equation}
here $\langle e_{IS}(T) \rangle$ is the solution to the saddle
point approximation to Eq.~(\ref{eq:partition}), and
$S_{conf}=k_B\ln(\Omega(e_{IS}))$ is the configurational entropy. 
$F_{vib}$, the intrabasin vibrational free energy,  is usually
written in the harmonic approximation as $F_{vib}=k_B T
\langle\sum_{i=1}^{M} \ln(\beta \hbar \omega_i(e_{IS})\rangle$ , where
$\omega_i(e_{IS})$ is the $i$-th normal mode frequency
($i=1...M$)) evaluated at the inherent structure, and
$\hbar$ is the Planck constant. The sum
of the logarithm of the normal modes frequencies describes the
volume (via the curvature) of the basin of attraction of the $IS$ 
in harmonic approximation.

Computer simulation results and theoretical insight~\cite{heuer00,sastry2001}
provide us valid models for the two crucial quantities
$\Omega(e_{IS})$ and $F_{vib}$, namely:
\begin{eqnarray}
\Omega(e_{IS}) de_{IS}&=&e^{\alpha N}
\frac{e^{-(e_{IS}-E_o)^2/2\sigma^2}}{\sqrt{2 \pi \sigma^2}}de_{IS}\label{eq:Omega}\\
F_{vib}(e_{IS},T) &=&k_B T[(a+be_{IS})-k_B T ln(\hbar \beta)]
.
\label{eq:fvib}
\end{eqnarray}

The hypothesis of a Gaussian landscape is supported by the
consideration that, if no correlation length diverges, 
$e_{IS}$ can be thought of as sum of the IS energy of several
independent subsystems. In this case the central limit theorem
suggests that, since the variance of the energy distribution in
each of these independent subsystem is finite, a gaussian
distribution will describe the distribution of
$e_{IS}$ values~\cite{heuer00}. We note that this hypothesis will
break down at the very low energy tail, where differences between
the gaussian distribution and the actual distribution become
relevant. The second hypothesis 
( $\sum_{i=1}^{M} \ln(\beta \hbar \omega_i(e_{IS}))= a + b e_{IS}$ )  
is not crucial, but it is supported by the results of numerical studies.

Substituting in Eq.~(\ref{eq:free_energy1}) and solving, one
obtains~\cite{sastry2001}:
\begin{eqnarray}
\langle e_{IS}(T)\rangle&=&( E_o - b \sigma^2 )-\sigma^2/k_BT
\label{eq:eis1}\\
- T S_{conf}(\langle e_{IS}(T) \rangle)&=&k_B T
\left(\frac{b^2\sigma^2}{2}-\alpha N\right)+b \sigma^2-\frac{\sigma^2}{2 k_B T}
\label{eq:Sconf1}\\
F_{vib} (T,\langle e_{IS}(T) \rangle)&=&
F_o(E_o,T)-k_B T b \sigma^2 (b+\beta).
\label{eq:fvib2}
\end{eqnarray}
\begin{figure}[t] 
\centering
%\vspace{1.cm}
\includegraphics[width=.8\textwidth]{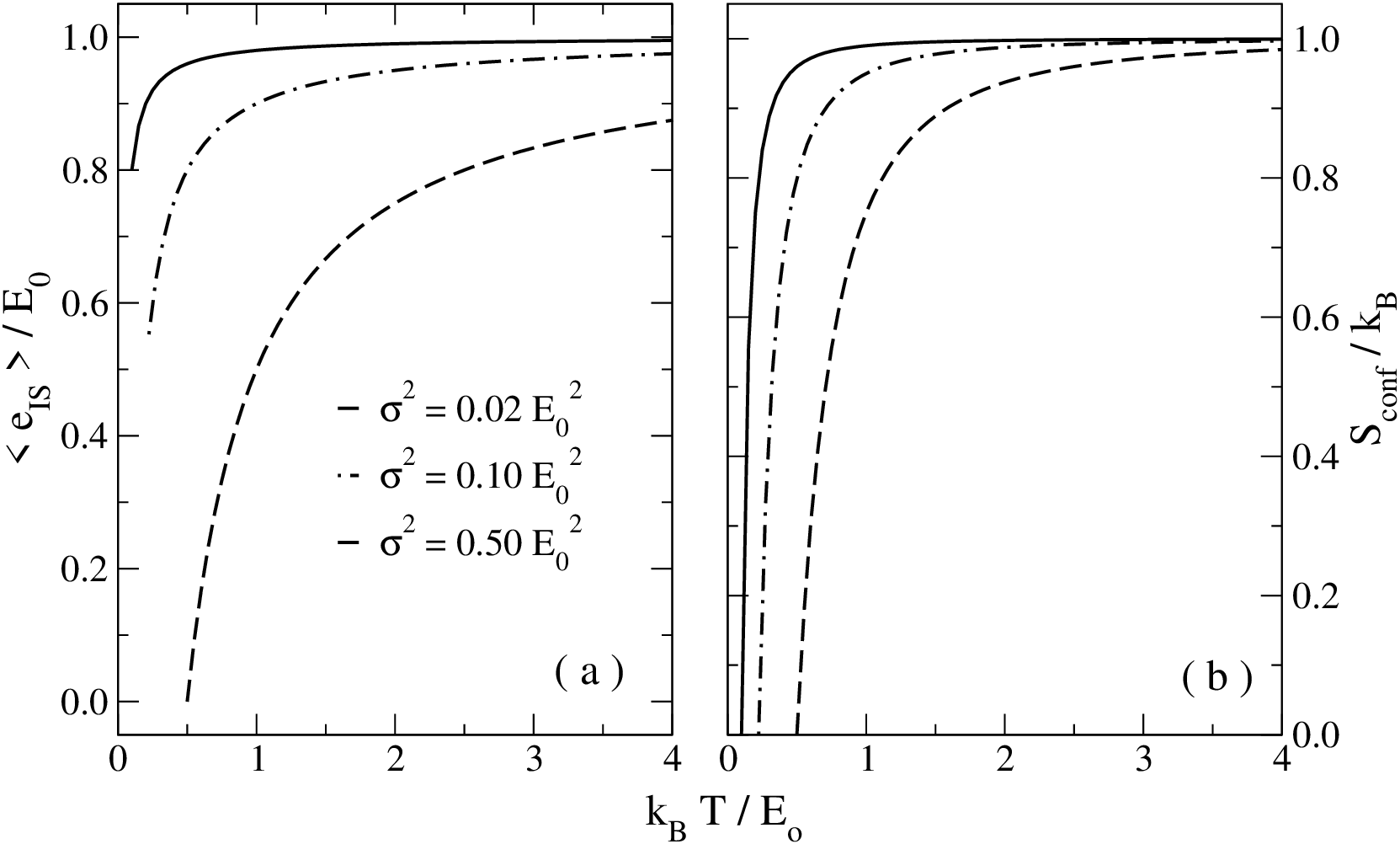}
\caption{$T$-dependence of $\langle e_{IS}\rangle$ {\it(a)} and 
$S_{conf}$ {\it(b)} in the approximation of 
Gaussian distribution of basin depths 
and $e_{IS}$-independence of the basin volume. 
Curves for $\alpha=1.0$ and three different values of $\sigma^2/E_0^2$ 
are shown. }
\label{fig:new}
\end{figure}
Therefore, $F(T,V)$ is expressed only in terms of proper
combinations of the parameters $\alpha$, $E_o$, $\sigma$, $a$, and
$b$ which are related to the statistical properties of the
PEL~\cite{lanave02} and to a particular relation between volume and depth of the basins. 
We also note that from a plot of $\langle e_{IS}\rangle$ vs $1/T$ it is possible
to evaluate $\sigma^2$ and $E_0$. A comparison between numerical
data and Eq.~(\ref{eq:Sconf1}) allows us to estimate $\alpha$.
In the case where all basins have the same volume ($b=0$),
Eqs.~(\ref{eq:eis1}) and (\ref{eq:Sconf1}) simplify considerably, and 
in terms of scaled quantities one obtains
\begin{eqnarray}
\langle e_{IS}(T)\rangle/E_o&=&1-
\frac{
 \sigma^2/E_o^2}{(k_BT/E_0)}
\label{eq:eis2}\\
S_{conf}(\langle e_{IS}(T) \rangle)/k_B&=&\alpha 
-\frac{\sigma^2/E_0^2}{2 (k_B T/E_0)^2}. \label{eq:Sconf2}
\end{eqnarray}
Within the Gaussian approximation, the lowest $e_{IS}$ value $e_K$
---characterized by $S_{conf}(e_K)=0$--- is the Kauzman energy
\begin{eqnarray}
\langle e_{K}(T_K)\rangle/E_o&=&1-\sqrt{ 2 \alpha \frac{\sigma^2}{E_o^2} },
\label{eq:ek}
\end{eqnarray}
and it is reached at a Kauzman temperature $T_K$ given by
\begin{eqnarray}
k_B T_k/E_o&=&\sqrt{ \frac {(\sigma^2/E_o^2)} {2\alpha}}.
\label{eq:tk}
\end{eqnarray}
The behavior of $\langle e_{IS}\rangle$ and $S_{conf}(T)$ as a function of $T$,
in reduced units, is shown in Fig.~\ref{fig:new}.
We note on passing that recent works by Speedy~\cite{speedyjpc} and 
Sastry~\cite{sastry2001} have attempted to correlate kinetic fragility 
to thermodynamic fragility~\cite{angellmartineznature}, 
suggesting that $\sigma$ and $\alpha$ are the statistical properties 
of the PEL which control the material fragility.
\begin{figure}[t]
\centering
%\vspace{1.cm}
\includegraphics[width=1.0\textwidth]{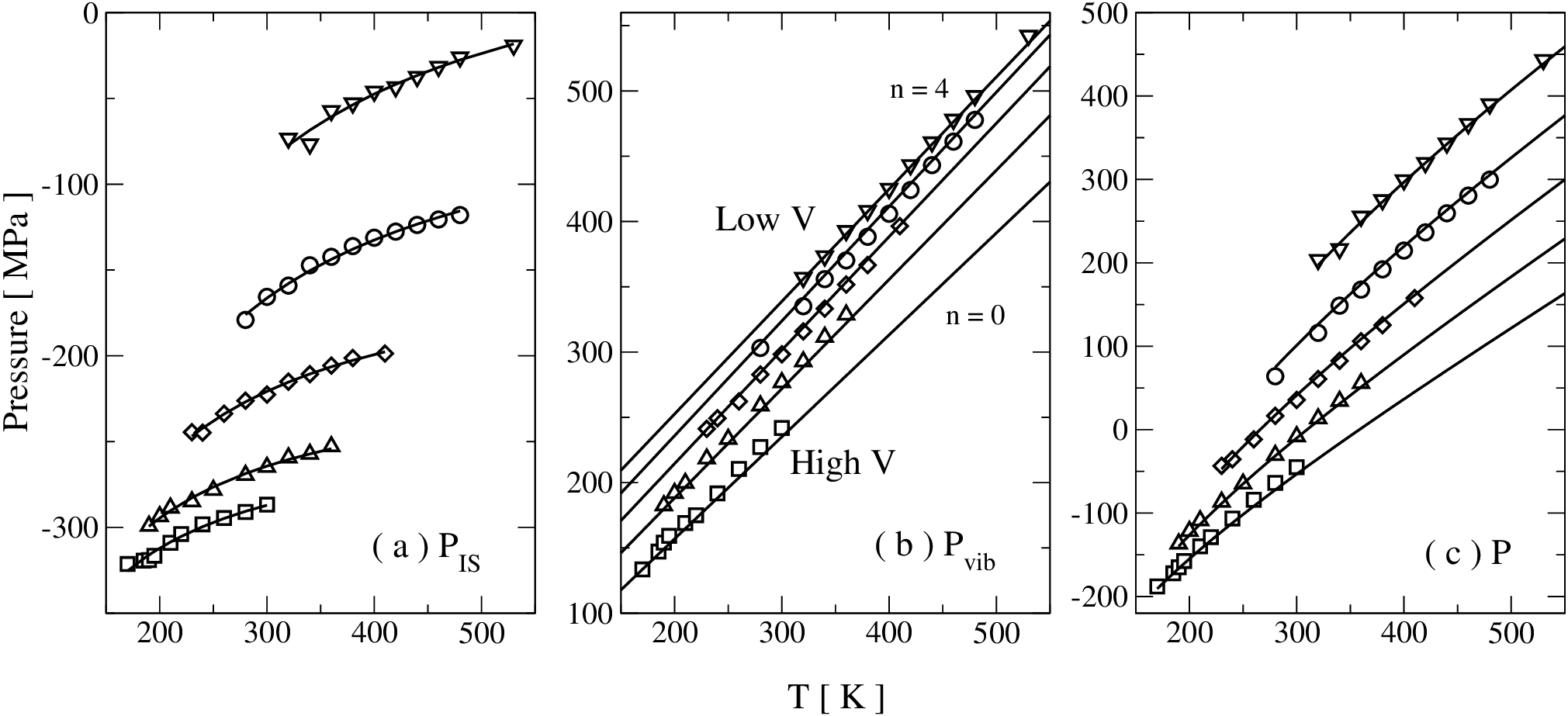}
\caption{
Comparison between the different contributions to the pressure
calculated according to the theory (solid lines) and by
MD simulations (symbols) for the LW model:
{\it(a)} Inherent structures contribution; {\it(b)} Vibrational contribution. The
curves have been shifted by $n\times 20 MPa$ to avoid overlaps;
{\it(c)} Total pressure. Details on the calculation of these
quantities can be found in Refs.~\protect\cite{lanave02,aging02}.}
\label{fig:equilibrium}
\end{figure}
\section{Equilibrium equation of state}
The generalization of Eq.~(\ref{eq:free_energy1}) to the volume
dependent case requires the determination of the volume dependence
of Eq.~(\ref{eq:Omega}), i.e., the formulation of an {\it ansatz}
for the joint probability ${\cal P}(e_{IS},V)$ to find a value of
$e_{IS}$ at a given volume $V$. We follow the equivalent approach
of fitting simultaneously the l.h.s. of
Eqs.~(\ref{eq:eis1})(\ref{eq:Sconf1}) and (\ref{eq:fvib})
determined by MD simulations at different volumes~\cite{lanave02}.
This procedure allows us to calculate directly the volume
dependence of the parameters $\alpha$, $E_o$, $\sigma$, $a$, and
$b$ introduced above. 

Substituting in Eq.~(\ref{eq:free_energy1}) we obtain $F(T,V)$ for
the model considered, and the equation of state can be finally
calculated via $P(T,V)=-\partial_V F(T,V)$ at $T$ constant.
From Eq.~(\ref{eq:free_energy1}) it is immediately clear that $P$
can be split into two contributions: a configurational part,
$P_{conf}$, related to the change in the number and depth of
available basins with $V$, and a vibrational part, $P_{vib}$,
related to the change in the volume of the basin with $V$ as
\begin{equation}
P(T,V,e_{IS})=P_{conf}(V,e_{IS})+P_{vib}(T,V,e_{IS}).
\label{eq:P_separ}
\end{equation}
Fig.~\ref{fig:equilibrium} shows the comparison among the MD
estimates of the different contributions to the pressure (symbols)
and the predictions of the above theory, for the case of the LW
model. The excellent agreement between the two sets of data
confirms the validity of the procedure introduced above which
provides us an effective equation of state for the system under
study based on the statistical properties of the landscape as
expressed in  Eqs.~(\ref{eq:Omega}) and (\ref{eq:fvib}).
\begin{figure}[t]
\centering
%\vspace{1.cm}
\includegraphics[width=.75\textwidth]{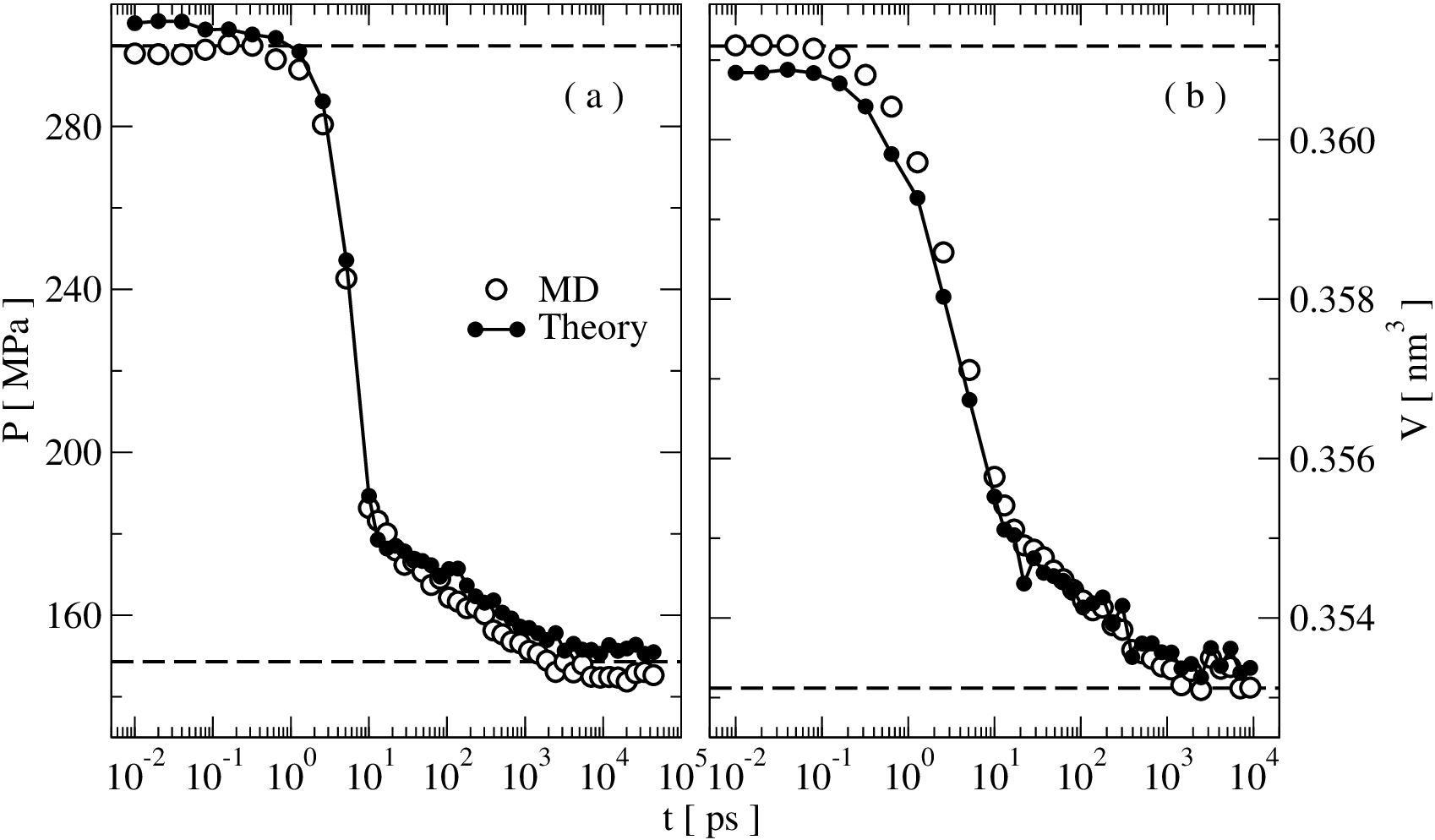}
\caption{OOE simulation protocols:
{\it (a)} Pressure evolution after a T-jump at constant volume
per molecule $V=0.345$ nm$^3$ from $480$ K to $340$ K;
{\it (b)} Volume evolution after a P-jump at constant temperature
$T=320$ K from $13.4$ MPa to $60.7$ MPa. Details can be found in 
Ref.~\protect\cite{aging02}.
}
\label{fig:quench-compress}
\end{figure}
\section{Out of equilibrium equation of state}
The possibility of a proper thermodynamical description of 
out-of-equilibrium systems has been widely
debated~\cite{davies,kurchan,teoneowanoezer,McKenna,angellmckenna}.
In particular it has been recognized that this should be possible
by adding one or more history-dependent parameters to the
equilibrium equation of state. The arguments discussed above allow
us to go further in this direction~\cite{aging02}. Indeed the only
hypothesis we made is the validity of Eqs.~(\ref{eq:Omega}) and
(\ref{eq:fvib}). In all the out-of-equilibrium conditions where
these two conditions are met, i.e., the system (gently) driven out
of equilibrium explores states which are typical at equilibrium,
the theory is expected to hold at the expenses of adding one
parameter to the equilibrium EOS. Looking at
Eqs.~(\ref{eq:eis1})(\ref{eq:Sconf1}) and (\ref{eq:fvib}), the
choice of the basin depth $e_{IS}$ as the additional parameter
turns out to be very natural.

To the extent of this extension the validity of
Eq.~(\ref{eq:P_separ}) in OOE conditions is crucial, allowing us
to link $P_{conf}$ and $P_{vib}$ to $e_{IS}$ and $V$. If this is
the case, the knowledge of $e_{IS}$ and $V$ is sufficient to
calculate both $P_{conf}$, $P_{vib}$ and their sum $P$ according
to Eq.~(\ref{eq:P_separ}). Similarly, the values of $P$, $T$ and
$e_{IS}$ are sufficient to predict $V$, since both
$P_{conf}(e_{IS},V)$ and $P_{vib}(e_{IS},T,V)$ can be estimated as
a function of $V$. The predicted $V$ is the value for which
$P_{conf}(e_{IS},V)+ P_{vib}(e_{IS},T,V)$ matches the external
(fixed) pressure.

In Fig.~\ref{fig:quench-compress} we show the comparison
among MD results and the predictions of the OOE equation of state
for two different OOE protocols via computer simulation.
In particular, we consider in Fig.~\ref{fig:quench-compress}(a)
the case of a T-jump at constant volume,
and in Fig.~\ref{fig:quench-compress}(b) a P-jump at
constant temperature. In the first case the dynamical evolution of
$e_{IS}$ together with the (fixed) values of $V$ and $T$
allow us to predict the dynamical evolution of $P$;
in the second one, the time dependence of $e_{IS}$ together
with $P$ and $T$ allow us to predict the evolution of $P$.
\begin{figure}[t]
\centering
%\vspace{1.cm}
\includegraphics[width=.80\textwidth]{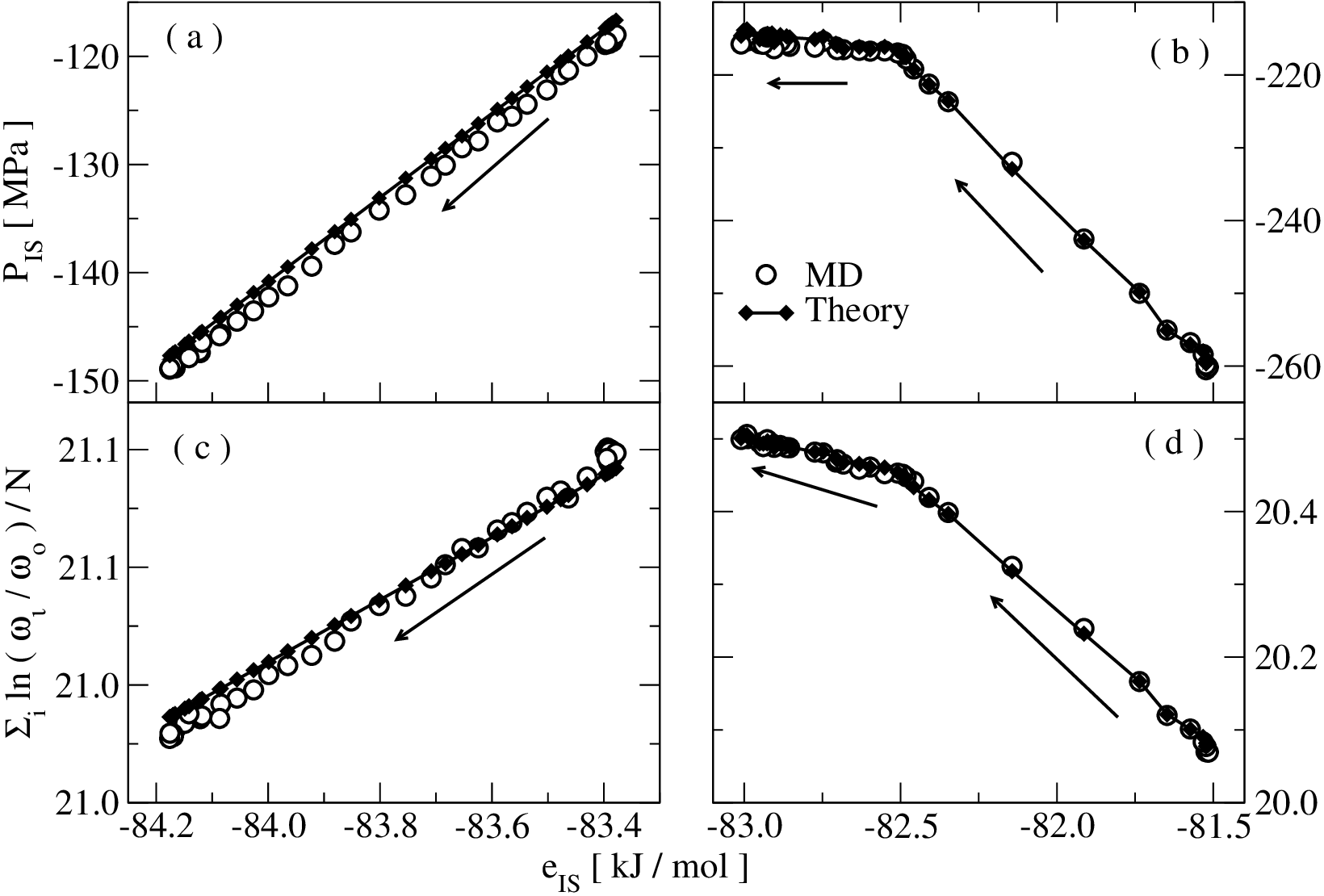}
\caption{Paths of the aging process in the $P_{IS}-e_{IS}$ plane
for the OOE protocols considered in Fig.~\protect\ref{fig:quench-compress}: 
{\it (a)} T-jump at constant pressure; {\it (b)} P-jump at
constant temperature. The arrows indicate the time evolution direction. 
Details can be found in Refs.~\protect\cite{lanave02,aging02}. Panels
(c) and (d) report the comparison between the "basin volume" during
the aging process with the "basin volume" of the corresponding
basin (same $e_{IS}$ and same $V$) explored in equilibrium
conditions. The basin volume is described by 
$\sum_{i=1}^{M} \ln( \omega_i(e_{IS}))$.}
\label{fig:PIS-eIS}
\end{figure}

An interesting representation of the aging processes discussed
above is the parametric plot in the $P_{IS}-e_{IS}$ plane. In
Fig.~\ref{fig:PIS-eIS} we show the path followed by the aging
system for the two protocols discussed above
(Figs.~\ref{fig:PIS-eIS}(a) and (b)). 
Panels (c) and (d) report the comparison between the 
basin volume described by the quantity 
$\sum_{i=1}^{M} \ln(\beta \hbar \omega_i(e_{IS}))$ during
the aging process and the basin volume of the corresponding
basin (same $e_{IS}$ and $V$) explored in equilibrium
conditions. In all cases the agreement
between the calculated quantities and the theoretical prediction
is quite good confirming the validity of our approach.

Fig.~\ref{fig:PIS-eIS}(b) is of particular interest,
showing that in the OOE dynamics following
a pressure jump can be recognized two different regimes.
For times shorter than the barostat time constant
(see Ref.~\cite{aging02} for details)
the system responses to the external increase of pressure
in a solid-like fashion, i.e., the PEL basins initially
populated are only deformed by the volume change.
Only for longer times, when the pressure has reached the
equilibrium value, the system starts to age among basins
different from the original ones.

\section{Conclusions}
In this paper we have reviewed some recent results on a general
approach to the thermodynamics of equilibrium supercooled liquids
and glasses~\cite{lanave02,aging02}. We have discussed how  it is
possible to formulate an equilibrium equation of state in terms of
quantities describing the statistical properties of the potential
energy landscape. These findings allow us to better understand the
nature of the different terms contributing to the total pressure
of the system, and fill the gap usually found among experiments
(usually performed at constant $P$) and computer simulations
(usually performed at constant $V$).

The generality of the hypothesis we have introduced, allow us to
generalize our approach to out of equilibrium conditions. In all
the cases where the introduced hypotheses are met, i.e., the
system ages among states typical at thermodynamical equilibrium,
it is possible to write down an OOE equation of state at the
expenses of the addition of one more parameter. This quantity can
be naturally chosen as the depth of the explored inherent
structures. The correctness of this generalization has been
checked under several OOE conditions. Its limits of validity,
under more severe OOE conditions where more then one additional
parameter is needed for a consistent description (like in the
so-called Kovacs memory experiments
~\cite{McKenna,angellmckenna,kovacs}) is currently under
investigation.  We foresee the possibility that, under large
variations of the temperature and/or pressure, different part of
the system will age with different speed producing, as a net
result, a material characterized by a distribution of $e_{IS}$ of
the composing subsystems different from the equilibrium one, and/or
a material for which the relation between basin volume and depth is
different from equilibrium.
%
%%%%%%%%%%%%
% REFERENCES
%%%%%%%%%%%%
%
\bigskip
\end{document}